\documentclass[pre,twocolumn,showpacs,amsmath,amssymb]{revtex4}
\newcommand{\EEE}{\mathbf{E}}

\newcommand{\rrr}{\mathbf{r}}
\newcommand{\nablabf}{\mathbf{\nabla}}

\newcommand{\0}{{\rm \scriptscriptstyle (0)}}
\newcommand{\1}{{\rm \scriptscriptstyle (1)}}

\usepackage{latexsym}

\usepackage{mathrsfs}

\usepackage{epsfig} 

\usepackage{dcolumn}  

\usepackage{bm}       

\DeclareFontFamily{OML}{eur}{\skewchar\font127} \DeclareFontShape{OML}{eur}{m}{n}{<5> <6> <7> <8> <9> gen * eurm <10><10.95><12><14.4><17.28><20.74><24.88>eurm10}{}
\DeclareSymbolFont{greek}{OML}{eur}{m}{n}

\DeclareMathSymbol{\upmu}{\mathord}{greek}{"16}

\begin{document}
\setlength{\unitlength}{1mm}

\author{Lars Rindorf}
\affiliation{COM$\bullet$DTU, Department of Communications, Optics
and Materials, Technical University of Denmark, DK-2800 Kgs.
Lyngby, Denmark}
\author{Niels Asger Mortensen}
\affiliation{MIC, Department of Micro and Nanotechnology, NanoDTU,
Technical University of Denmark, DK-2800 Kgs. Lyngby, Denmark}
\title{Calculation of optical-waveguide grating characteristics using Green's functions and the Dyson's equation}

\begin{abstract}
We present a method for calculating the transmission spectra,
dispersion, and time delay characteristics of optical-waveguide
gratings based on Green's functions and Dyson's equation. Starting
from the wave equation for transverse electric modes we show that
the method can solve exactly both the problems of coupling of
counter-propagating waves (Bragg gratings) and co-propagating
waves (long-period gratings). In both cases the method applies for
gratings with arbitrary dielectric modulation, including all kinds
of chirp and apodisation and possibly also imperfections in the
dielectric modulation profile of the grating. Numerically, the
method scales as $\mathcal{O}(N)$ where $N$ is the number of
points used to discretize the grating along the propagation axis.
We consider optical fiber gratings although the method applies to
all 1D optical waveguide gratings including high-index contrast
gratings and 1D photonic crystals.
\end{abstract}
\maketitle

\section{Introduction}

Spectral properties of optical fiber waveguide
gratings~\cite{meltz1989, hill1997} have typically been addressed
with coupled-mode theory
(CMT)~\cite{kogelnik1972,yariv1973,Erdogan:1997,Erdogan:1997IEEE}.
The theory relies on entities with a clear physical interpretation
and despite its approximations (slowly varying amplitude, small
index modulation, and synchronous approximations) it has proved
successful in accounting for the behavior of long-period gratings
(LPGs) and Bragg gratings (BGs) with uniform dielectric
modulation, i.e. a dielectric modulation of the grating with
constant period and peak value. Over the years corrections to CMT
for non-uniform gratings as well as more general solutions have
been addressed in a number of papers including
Refs.~\onlinecite{sipe1994,sullivan1995,poladian1996,Passaro:2002,Passaro:2002a,wellerbrophy1985}.

Others have addressed the problem of gratings with arbitrary
dielectric modulations through iterative solution of integral
equations in inverse scattering methods, such as the
Gel'fand--Levitan--Marchenko coupled
equations~\cite{peral1996a,peral1996b}. The iterative solution of
the integral equations scales as $\mathcal{O}(N^3)$, where $N$ is
the number of points along the propagation axis, and does not
always converge for strong Bragg gratings. The convergence problem
can be remedied by using the approximate layer-peeling
algorithm~\cite{feced1999} which is based on a combination of
transfer matrices~\cite{yamada1987} and a differential scattering
formulation~\cite{bruckstein1985}. Originating in geophysics the
layer-peeling algorithm divides a non uniform grating into a
number of uniform segments. Numerically the method scales as
$\mathcal{O}(N^2)$.

In this paper we offer a Green's function method (GFM) based on
the solution of the \emph{Dyson}'s integral equation that is exact
for coupling of co-propagating transverse electric waves as well
as counter-propagating transverse electric waves. The numerical
implementation scales as $\mathcal{O}(N)$. Our calculation of the
Green's function formally relies on methods from quantum mechanics
and the many-body theory of condensed matter systems, see e.g.
Refs.~\onlinecite{fetterwalecka, Bruus}. Interpreting the Green's
function as a space(-time) propagator immediately provides us a
link to the transmission and reflection properties of
finite-length gratings and thus forms a physical interpretation of
the Green's function.

The GFM thus offer an exact integral method, which also may be
formulated in terms of transfer matrices. Furthermore,
perturbative calculations may be readily be carried out with the
Dyson's equation. To demonstrate the method we calculate spectra,
time delay, and dispersion characteristics for uniform, as well as
apodised and chirped fiber gratings.

The Green's functions and the Dyson's equation have previously
been used in electromagnetical systems for a variety of problems
(see Ref.~\onlinecite{martin1995} and references therein), among
those finite size 2D photonic
crystals~\cite{martin1999,rahachou2005,zhao2005}.

The paper is organised as following. In
Sec.~\ref{opticalwaveguide} we describe the problem of optical
waveguide gratings. In Sec.~\ref{sec:formalism} we describe the
problem in terms Maxwells equations. In
Secs.~\ref{sec:greensfunctions},\ref{theintermediategreensfunction}
we introduce the Green's function formalism and the Dyson's
equation. In Sec.~\ref{longperiodgrating} we apply the theory to
long-period gratings and in Sec.~\ref{bragggratings} to Bragg
gratings. In Secs.~\ref{discussion} and \ref{conclusion}
discussions and conclusions are provided.

\section{Optical waveguide gratings}
\label{opticalwaveguide}
The waveguide grating is a scattering region introduced in the
waveguide and is effectively equivalent to a one dimensional finite size photonic
crystal. Waveguides support a number of guided modes; one or two
fundamental modes and an infinite number of higher order modes.
All modes have a forward and a backward propagating part. In the
so-called \emph{long-period grating} different forward propagating
modes couple at a resonant frequency. Forward scattering involves
a very small momentum transfer $\delta\beta=k\Delta n^{\rm eff}$
where $k=\omega/c=2\pi/\lambda$ is the free-space wavenumber and
$\Delta n^{\rm eff}$ is the difference in effective indices of the
resonant modes involved. Since $\delta\beta\Lambda_G\sim 1$ at
resonance, the period $\Lambda_G$ of such a grating is necessarily
long, hence its name. In Bragg gratings the forward and backward
propagating parts of the same mode couple involving a large
momentum transfer $\delta\beta=2\beta=2kn^{\rm eff}$ and thus
Bragg gratings necessarily have a short grating period
$\Lambda_G$.

\begin{figure}[h]
\begin{center}
\includegraphics[angle = 0, width = 0.5\textwidth]{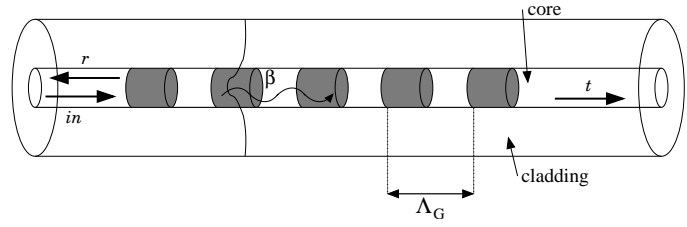} \caption{An optical fiber
waveguide with a grating consisting of a  dielectric modulation in
the core (grey area) with period $\Lambda_{G}$. The incident
electric mode $in$ with propagation constant $\beta$ is partially
reflected, ($r$), and partially transmitted ($t$) by the grating.
No radial symmetry is assumed in the presented theory.}
\label{fig:opticalwaveguide}
\end{center}
\end{figure}

The dielectric modulation seen in Fig.~\ref{fig:opticalwaveguide}
can be perfectly periodic with $\Lambda_\textrm{G}$ (uniform
grating) or can be quasi-periodic (non-uniform grating). It is
possible to alter the spectral characteristics of a grating by
using \emph{apodisation} and \emph{chirp} in the dielectric
modulation Ref.~\cite{cross1977}. Such techniques have made
gratings widely used as highly selective bandpass filters
\cite{ibsen1998}, dispersion compensators and pulse
compressors~\cite{ouellete1987, roman1993} and optical
sensors~\cite{lee2003}. With apodisation the magnitude of the peak
dielectric modulation varies across the length of the grating.

With chirp the period of the grating varies across the grating.
Apodisation and chirp, however, may also appear as an unwanted
error in an intended uniform grating due to  fabrication imperfections.

Side-lobe suppression can be done by an apodisation known as
raised-Gaussian apodisation~\cite{deyerl2004} where the dielectric
modulation has a Gaussian envelope.

In a linear chirp the grating period varies as $\Lambda_G(z) =
\Lambda _G \big(1 + \frac{c_1}{2}\frac{z}{L}\big)$ where $c_1$ is
the chirp parameter and $L$ is the length of the grating.
Apodisation can be visualised as segments of uniform gratings that
are concatenated. The spectra are different due to different peak
dielectric modulation, but share the same resonance wavelength. A
chirped grating can be visualised as the concatenation of uniform
grating segments with different grating periods.

\section{General formalism}
\label{sec:formalism}
In the following we consider temporal harmonic solutions to
Maxwell's equations with electrical fields of the form $
\EEE(\rrr,t) = \EEE_\omega(\rrr)e^{-i\omega t}, $ with the
subscript $\omega$ indicating the frequency dependence. The
electrical field is a solution to a vectorial wave equation, which
has the form of a generalised eigenvalue
problem~\cite{joannopoulos,martin1995},
\begin{equation}
\nablabf \times \nablabf \times \EEE_\omega(\rrr) -
k^2\boldsymbol{\varepsilon}_r(\rrr_\bot)\EEE_\omega(\rrr) -
k^2\boldsymbol{\varepsilon}_G(\rrr)\EEE_\omega(\rrr) =0,
\label{eq:eigenfreq_r}
\end{equation}
where the dielectric functions, $\boldsymbol{\varepsilon}$, are
$3\times3$ dielectric tensors. $\rrr_\bot$ is the coordinate
vector orthogonal to the axis of the fiber which we define to be
parallel with the z-axis. The subscript `$r$' refers to what we
will call the reference system, which is the bare fiber without
any grating. $\boldsymbol{\varepsilon}_r$ is thus translational
invariant along the z-axis. The subscript `$G$' refers to the
changes in the dielectric function induced by the grating. The
grating begins at $z_i$ and ends at $z_f$ and
$\boldsymbol{\varepsilon}_G(\rrr)$ is thus zero outside this
interval. Usually $\boldsymbol{\varepsilon}_G(\rrr)$ is denoted as
$\delta\boldsymbol{\varepsilon}_G(\rrr)$ because it is assumed
that it is small, but here we use the former notation to emphasise
that the presented theory is formally valid for any dielectric
profile including high-index contrast profiles since
Eq.~(\ref{eq:eigenfreq_r}) is completely general. In the following
$\boldsymbol{\varepsilon}_G(\rrr)$ is referred to as the
dielectric modulation and we will for simplicity omit the
subscript on $\EEE_\omega$.

In order to solve Eq.~(\ref{eq:eigenfreq_r}) for a general
$\boldsymbol{\varepsilon}_G$ we will use a Hilbert function space
spanned by the monochromatic (fixed $k$) eigensolutions to
to the reference system in Eq.~(\ref{eq:eigenfreq_r})
\begin{equation}
\EEE_{m}^{\0}(\rrr) =\EEE_{m}^{\0}( \rrr _\bot )e^{i\beta _m z},
\end{equation}
where $\beta_m$ is the propagation constant of the mode $m$. For
the inner product we have $ \big< E_m^{\0} |  E_n^{\0} \big> =
\int d\rrr\, \EEE_m^{\0\dagger}(\rrr)\cdot \EEE_n^{\0}(\rrr)$,
where $\dagger$ is the Hermitian conjugate (transpose and complex
conjugate) of the field vector. We choose the eigenmode basis
$\big\{E_m^{\0}\big\}_{m=1}^M$ to be orthonormal, $\big< E_m^{\0}|
\hat{\varepsilon}_r| E_n^{\0} \big> =
\big(\frac{\beta_m}{k}\big)^2\delta_{mn}$,
where 
 $\delta$ is the Kronecker delta function.
This orthonormalisation follows from the fact that
Eq.~(\ref{eq:eigenfreq_r}) is a generalised eigenvalue problem.
 Any linear operator $\hat{O}$ on a Hilbert space of finite dimension can be
represented by a matrix  acting on a state vector as $\hat{O} = \sum_{mn} O_{mn}
|E_m^{\0}\big>\big<E_n^{\0}|$, where $\hat{O}$ has matrix elements
$O_{mn}  = \int d\rrr\, \EEE_m^{\0 \dagger}(\rrr) \mathbf{O}(\rrr)
\EEE_n^{\0}(\rrr)$.

For transverse electric modes, dielectric tensors with
$\varepsilon_{xz}=\varepsilon_{zx}=\varepsilon_{yz}=\varepsilon_{zy}=0$,
and it is possible to integrate out the transverse degrees of
freedom in Eq.~(\ref{eq:eigenfreq_r}) to obtain
\begin{equation}
\frac{d^2}{dz^2} \big|E\big> + k^2\hat{\varepsilon}_r \big|E\big>
+ k^2\hat{\varepsilon}_G(z) \big|E\big> =0, \label{eq:eigenfreq}
\end{equation}
where $\big|E\big>$ is a linear combination of eigenmodes with
scalar coefficients, i.e. (state) vector in the eigenmode basis
$\big\{E_m^{\0}\big\}_{m=1}^M$. We have thus transformed the fully
vectorial eigenvalue problem of Eq.~(\ref{eq:eigenfreq_r}) into a
scalar eigenvalue problem given by Eq.~(\ref{eq:eigenfreq}) in the
Hilbert function space spanned by the fully vectorial eigenmodes.

\section{Green's Functions}
\label{sec:greensfunctions}
In the following we address Eq.~(\ref{eq:eigenfreq}) with
the aid of Green's functions which will in turn provide us with
insight in the transmission and reflection properties of the
grating.

\subsection{Solution with Dyson's equation}
Let $G^{\0}$ be the zero-order Green's function solution to the
eigenvalue equation, Eq.~(\ref{eq:eigenfreq}) with
$\varepsilon_G=0$. This has the analytical solution
\begin{equation}
iG_{mn}^{\0}(z, z') = \delta_{mn} \frac{e^{i\beta _m|z-z'|}}{2\beta_m}.
\end{equation}
 The zero-order Green's function
only depends on $z'-z$ in agreement with the assumption of
translational invariance of the reference system. The zero-order
Green's function can conveniently be split into a spatial forward
propagating and a backward  propagation part
\begin{eqnarray}
\theta(z-z')
iG_{ \textrm{+},mm}^{\0}(z, z') +
\theta(z'-z)
iG_{ \textrm{-},mm}^{\0}(z, z')\nonumber \\
=
\theta(z-z') \frac{e^{i\beta _m
(z-z')}}{2\beta_m} +
\theta(z'-z)
 \frac{e^{-i\beta _m
(z-z')}}{2\beta_m}.  \label{eq:g0}
\end{eqnarray}
The subscripts `+' and `-' refer to the forward and backward
propagating part of the Green's function, respectively, and
$\theta$ is Heaviside's stepfunction. We also note that $G^{\0}_+$
and $G^{\0}_-$ have the convenient mathematical property that they
each split in their variables: $e^{\pm i\beta _m (z-z')}=e^{\pm
i\beta_mz}e^{\mp i\beta_mz'}$. In terms of $G^{\0} $ the Green's
function solution to the full grating problem
Eq.~(\ref{eq:eigenfreq}) is given by the \emph{Dyson's
equation}~\cite{martin1995}
\begin{multline}
i\boldsymbol{G}(z,z') = i\boldsymbol{G}^{\0}(z,z') \\
 -i \int_{z_i}^{z_f} \hspace{-0.25cm} dz'' \,i\boldsymbol{G}^{\0}(z,z'' )
k^2\hat{\varepsilon}_{G}(z'') i\boldsymbol{G}(z'',z' ).
\label{eq:dysonsequation}
\end{multline}
The integration interval is changed to $[z_i,z_f]$ since
$\varepsilon_{G}$ is zero outside this interval.

\subsection{Physical properties of the Green's functions}
\label{sec:greenstrans}
The transmission and reflections properties may be described in a
scattering basis formulation where ingoing and outgoing modes are
related to each other by a scattering matrix
\begin{equation}
{\boldsymbol S}=\begin{pmatrix} {\boldsymbol r} & {\boldsymbol
t}'\\{\boldsymbol t} & {\boldsymbol r}'
\end{pmatrix}.
\end{equation}
Here, $t_{mn}$ is the transmission amplitude for forward
scattering (left-to-right direction) from mode $n$ to mode $m$ and
$r_{mn}$ is the reflection amplitude for back-scattering
(left-to-left direction) from mode $n$ to mode $m$. Likewise, $t'$
and $r'$ are amplitudes for the opposite directions. We restrict
ourselves to lossless media for which $\boldsymbol S$ is unitary
and symmetric so that ${\boldsymbol t}' = -{\boldsymbol
t}^\dagger$, ${\boldsymbol r}' = {\boldsymbol r}^\dagger$, and
${\boldsymbol 1} = {\boldsymbol r}^\dagger {\boldsymbol r} +
{\boldsymbol t}^\dagger {\boldsymbol t}.$ 
 The Green's function is naturally interpreted as a propagator for
field amplitudes and the full Green's function may be related to
the bare Green's function via~\cite{Bruus}
\begin{subequations}
\begin{eqnarray}
i G_{mn}(z_f, z_i) &=&   t_{mn} \frac{e^{i(\beta_m z_f - \beta_n z_i)}}{2\sqrt{\beta_m\beta_n}}  \\
i G_{mn}(z_i, z_i) &=&  r_{mn} \frac{e^{i(\beta_m - \beta_n)
z_i}}{2\sqrt{\beta_m\beta_n}}
\end{eqnarray}
\end{subequations}
and the transmission and reflection probabilities become
\begin{subequations}
\begin{align}
T_{mn} = |t_{mn}|^2= 2^2\beta_m\beta_n |iG_{mn}(z_f,z_i)|^2,
\label{eq:Tmn}\\
R_{mn} = |r_{mn}|^2=2^2\beta_m\beta_n |iG_{mn}(z_i,z_i)|^2.
\label{eq:Rmn}
\end{align}
\end{subequations}
The group delays of the transmission and reflection are given by
the phase of the amplitude coefficients, $t_{mn}$ and $r_{mn}$. The group delay for the
scattering $n\to m$ is given by
\begin{eqnarray}
\tau _{mn} = \arg\big( t_{mn} \big)
, \label{eq:groupdelay}
\end{eqnarray}
where $\arg(x)$ is the argument of the complex number of $x$. For
the group delay of the reflection scattering similar arguments
hold and the group delay may be obtained by substituting $t_{mn}$
with $r_{mn}$ in Eq.~(\ref{eq:groupdelay}). The dispersion, $D$,
of the $n\to m$ scattering is defined as the derivative of the
group delay with respect to wavelength
\begin{eqnarray}
D _{mn} \equiv \frac{d \tau _{mn}}{d \lambda}
.\label{eq:dispersion}
\end{eqnarray}

\section{The intermediate Green's function}
\label{theintermediategreensfunction}
The dielectric modulation of the grating not only causes the
different modes to coupled, but also the individual modes to
change. This phenomenon is also refereed to as
\emph{self-coupling} and has the physical interpretation that the
grating causes the individual propagation constant of a mode
($\beta_m=n_m^{\rm eff} k$) to change slightly. Self-coupling is
typically of second importance to cross-mode coupling in fiber
gratings and it can in many instances be ignored. However, we can
at no cost incorporate it in the free forward and backward Green's
function propagators and we refer to these new propagators
$\tilde{G}_\pm$ as the \emph{intermediate Green's functions}. By
this trick we eventually end up with a Dyson's equation for the
full Green's function with a particular simple structure. To see
this we split $\hat{\varepsilon}_{G}$ into two Hermitian
operators: a self-coupling operator (s), which is diagonal in its
indices, and a cross (x) coupling operator which is purely
off-diagonal:
\begin{subequations}
\begin{align}
\hat{\varepsilon}_{G}(z) &= \hat{\varepsilon}^{\rm s}(z) + \hat{\varepsilon}^{\rm x}(z)\\
\varepsilon^{\rm s}_{mn}(z) &\equiv \delta_{mn}\varepsilon_{G,mm}(z)\\
\varepsilon_{mn}^{\rm x}(z) &\equiv \varepsilon_{G,mn}(z) -
\varepsilon^{\rm s}_{mn}(z)
\end{align}
\end{subequations}
The diagonal, self-coupling component, $\varepsilon^\textrm{s}$,
is often refereed to as the `DC' component and the cross coupling
component, $\varepsilon^\textrm{x}$, as the `AC' component in
analogy with electric engineering. We first consider the forward
Green's function. To zero order in the self-coupling the
intermediate Green's function equals the free Green's function,
$\tilde{G}^{\0}_{\rm +}(z',z)=G^{\0}_{\rm +}(z',z)$. Iterating
Dyson's equation and keeping terms to linear order in $\hat
\varepsilon^\textrm{s}$ gives the corresponding contribution to
the intermediate Green's function
\begin{align}
-&i\int_{z} ^{z'}dz'' \,
i{\boldsymbol G}^{\0}_{\rm +}(z',z'') k^2{\hat{\varepsilon}}^{\rm s}(z'')
i{\boldsymbol G}^{\0}_{\rm +}(z'',z)
\nonumber \\
&= i\boldsymbol{G}^{\0}_{{\rm +}}(z',z)\frac{-i
k^2}{2\boldsymbol{\beta}}\int_{z} ^{z'}dz''\, \hat{\varepsilon}^\textrm{s}(z'').
\end{align}
With the operator
\begin{equation}
\Delta\boldsymbol{\beta}(z',z)\cdot(z'-z) \equiv
\frac{k^2}{2\boldsymbol{\beta}}\int_{z} ^{z'}dz''
\hat{\varepsilon}^\textrm{s}(z'') , \label{eq:deltabeta}
\end{equation}
we may solve Dyson's equation, Eq.~(\ref{eq:dysonsequation}),
straightforwardly to any order. One might wonder why we chose to
define a function as in Eq.~(\ref{eq:deltabeta}). This definition
has the advantages that the left hand side is distributive in $z$
and $z'$, $\Delta\boldsymbol{\beta}(z',z)\cdot(z'-z) =
\Delta\boldsymbol{\beta}(z',z'')\cdot(z'-z'') +
\Delta\boldsymbol{\beta}(z'',z)\cdot(z''-z)$, and if the equation
is divided by $z'-z$, then it can be seen that $\Delta\beta_{mm}$,
is proportional to an average of $\boldsymbol\varepsilon_G$ over
$z'-z$. We will use these properties in the following. For the
backward propagating intermediate Green's function the analysis
above may be repeated. Using the distributive property of
$\Delta\boldsymbol{\beta}$ and introducing
\begin{equation}
\tilde{\boldsymbol \beta}(z)z \equiv {\boldsymbol \beta}z +
{\Delta\boldsymbol {\beta}}(z,z_i)\cdot(z-z_i)
\label{tildebeta}
\end{equation}
we may write the forward and backward propagating intermediate Green's functions as
\begin{subequations}
\begin{align}
\theta(z-z')i\tilde{\boldsymbol{G}}_+(z,z')
=
\theta(z-z')\frac{e^{i\tilde{\boldsymbol{\beta}}(z) z}
e^{-i\tilde{\boldsymbol{\beta}}(z')z'}}
{2\boldsymbol{\beta}},
\label{eq:intermediategreensfunction+}\\
\theta(z'-z)i\tilde{\boldsymbol{G}}_-(z,z')
=
\theta(z'-z)
\frac{e^{-i\tilde{\boldsymbol{\beta}}(z) z}
e^{i\tilde{\boldsymbol{\beta}}(z')z'}}
{2\boldsymbol{\beta}},
\label{eq:intermediategreensfunction-}
\end{align}
\end{subequations}
which are diagonal matrices in their mode indices. The Dyson
equation (\ref{eq:dysonsequation})  may be rewritten in terms of
the intermediate Green's functions and the cross coupling operator
\begin{align}
i\boldsymbol G(z',z) &=  i\tilde{\boldsymbol G}(z',z)
\nonumber \\&
-i\int_{z_i} ^{z_f}dz'' \,
i\tilde{\boldsymbol G}(z',z'') k^2{\hat{\varepsilon}}^{\rm x}(z'')
i{\boldsymbol G}(z'',z)
.\label{eq:dysontilde}
\end{align}

\subsection{Numerical evaluation of Dyson's equation}
\label{sec:numeval}

The Dyson's equation (\ref{eq:dysontilde}) describes a system of $M
\times M$ coupled equations. Due to the fact that the matrix
$i\tilde{\boldsymbol{G}}$ is diagonal in its mode indices the
different matrix columns of $i\boldsymbol{G}$ are not coupled to
each other, and the Dyson's equation constitute $M$ problems of
size $M$ instead of a single problem of size $M\times M$. If only
a single electric mode is incident on the grating (as is often,
but not always, the case) only a single column of the matrix
Dyson's equation needs to be solved. For large $M$ and $N$ this gives
a significant reduction in time and memory consumption.

The matrix Dyson's equation (\ref{eq:dysontilde}) may be solved by
simple iteration by inserting $\tilde{\boldsymbol{G}}$ on the
position of $\boldsymbol{G}$ on the right hand side of the
equation to obtain $\sum_{\alpha = 0} ^1 i\boldsymbol{G}^{ \rm
\scriptscriptstyle (\alpha)} (z,z')$ on the left hand side. This
in turn may be inserted  to find $\sum_{\alpha = 0} ^2
i\boldsymbol{G}^{ \rm \scriptscriptstyle (\alpha)} (z,z')$. Such
an iteration may be continued until any order. In this work we
evaluated Dyson's equation with the discrete dipole approximation
\cite{draine1994}. In the approximation the scattering interval
$[z_i,z_f]$ is discretised into $N$ cells. Each cell is assumed
small enough that the variations of the electric field $\EEE (z)$
and the dielectric modulation $\boldsymbol \varepsilon_G (z)$ are
negligible within the cell.

The numerical task of evaluating a single iteration with
Eq.~(\ref{eq:dysontilde}) generally scales as $N^2$ where $N$ is
the number of cells on the discretised $z$ grid. However due to
the distributive property of the forward and backward propagating
intermediate Green's functions shown in
Eqs.~(\ref{eq:intermediategreensfunction+},\ref{eq:intermediategreensfunction-}),
the calculation scales as $N$.

\section{Copropagating waves - Long-period gratings}
\label{longperiodgrating}

In a LPG the power of an incident core mode is
coupled into a forward propagating cladding mode which has a
different, smaller propagation constant, $\beta$, but the same
frequency $\omega$. The cladding modes are slightly lossy causing
the mode power of the cladding mode to slowly dissipate into the
environment of the waveguide. Thus in practice the total energy
transmitted through the waveguide is reduced
 by the amount of energy transmitted into the cladding mode.
Physically, the cladding mode losses are small over the length of
a grating and may be neglected in the calculations.

\subsection{Resonance condition}
\label{sec:res_lpg}
To analyse the resonance, we may neglect higher order
contributions than first order in $\hat \varepsilon^{\rm x}$ (but
retain all orders in $\hat \varepsilon ^{\rm s} $) to the total
Green's function
\begin{align}
i\boldsymbol{G}(z_f,z_i) &= i{{\boldsymbol{G}}}^{\0}(z_f,z_i) + i{{\boldsymbol{G}}}^{\1}(z_f,z_i)\nonumber\\
 &= \frac{e^{-i\tilde{\boldsymbol{\beta}}(z_f) z_f}} {\sqrt{2\boldsymbol{\beta}}}\Big(
\boldsymbol{1} - i  \boldsymbol{\kappa}(z_f,z_i)(z_f-z_i)
 \Big) \frac{e^{i\tilde{\boldsymbol{\beta}}(z_i)  z_i}} {\sqrt{2\boldsymbol{\beta}}},
\end{align}
where the cross-coupling function, $\boldsymbol{\kappa}$, is defined as
\begin{align}
\kappa_{mn}(z',z)\cdot(z'-z) = & \nonumber \\
\frac{k^2}{2\sqrt{\beta_m\beta_n}} \int_{z}
^{z'}\hspace{-0.25cm} dz'' \hspace{-0cm} &e^{-i\big(\tilde{\beta}
_m(z'') -\tilde{\beta}_n(z'')\big)z''} \varepsilon^x_{mn}(z'')
\label{eq:kappa_LPG}.
\end{align}
We can thus estimate the resonant wavelength by finding the
maximum of $|\kappa_{mn}|$ with respect to the wavelength. For
uniform gratings with constant period and strength we may
rearrange the integrand of the expression as\footnote{Here we have
defined the average of a function, $f$, as $\overline{f(z)}\equiv
\frac{1}{z_f-z_i}\int^{z_f}_{z_i}dz f(z)$}
\begin{align}
e^{-i\big(\tilde{\beta}_m(z) -\tilde{\beta}_n(z)\big)z}& \varepsilon^x_{mn}(z)
= e^{-i\big(\overline{\tilde{\beta}_m(z)}
 -\overline{\tilde{\beta}_n(z)}\big)z} \times
\nonumber \\
&e^{-i\big(
\tilde{\beta}_m(z)-\overline{\tilde{\beta}_m(z)}
 -\tilde{\beta}_n(z)+\overline{\tilde{\beta}_n(z)}
 \big)z} \varepsilon^x_{mn}(z).
\label{eq:lpgcondunifmellemresultat}
\end{align}
The middle factor on the right hand side of Eq.
(\ref{eq:lpgcondunifmellemresultat}) does not have a different
period than $\varepsilon^x_{mn}(z)$. The self-coupling thus
changes the resonant frequency from $8\beta_m -\beta_n =
\frac{2\pi}{\Lambda_G}$ to an average $\overline {\beta_m(z)} -
\overline{\beta_n(z)} = \frac{2\pi}{\Lambda_G}$ in the limit
$\frac{z_f-z_i}{\Lambda_G}\to \infty$. This result is the same as
that of the \emph{synchronous approximation} used in
CMT~\cite{Erdogan:1997,Erdogan:1997IEEE} which arises from
different arguments.

\subsection{Numerical solution of Dyson's equation}
\begin{figure}[h]
\begin{center}
\includegraphics[angle = 0, width = 0.5\textwidth]{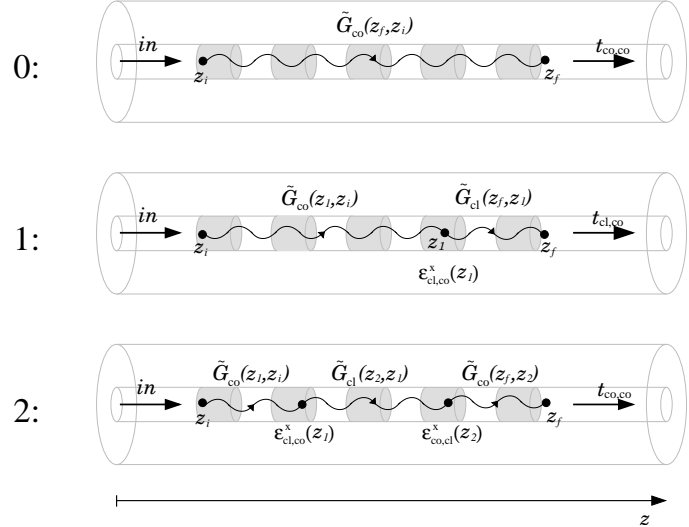} \caption{Diagrammatic representation
the zero to second order contributions to the total Green's
function Eq.~(\ref{eq:dyson_num_lpg}). An incident core mode
interacts with a copropagating cladding mode through the
dielectric modulation. } \label{fig:lpg}
\end{center}
\end{figure}

We consider a system of a pair of coupling modes: a core mode and
a cladding mode. Only the forward propagating Green's functions
need to be considered as the coupling to the backward propagating
modes is far from resonance and is negliable. The use of only
forward propagating Green's functions alters the integration
limits in the Dyson's equation, Eq.~(\ref{eq:dysontilde}), due to
the step function in the forward propagating intermediate Green's
function, Eq.~(\ref{eq:intermediategreensfunction+}). We consider a
system where only a single core mode is incident on the grating
and thus need only to solve the first column of the matrix Dyson's
equation, Eq.~(\ref{eq:dysontilde})
\begin{widetext}
\begin{align}
\left[
\begin{array}{cc}
iG_{\rm co,co}(z,z_i)   \\
iG_{\rm cl,co}(z,z_i)
\end{array}
\right]
&=
\left[
\begin{array}{cc}
i\tilde{G}_{\rm co}(z,z_i)  \\
0
\end{array}
\right]
-ik^2
\int_{z_i}^{z}\hspace{-0.25cm}dz'
\left[
\begin{array}{cc}
i\tilde{G}_{\rm co}(z,z') & 0 \\
0 & i\tilde{G}_{\rm cl}(z,z')
\end{array}
\right]
\left[
\begin{array}{cc}
0 &\varepsilon_{\rm co,cl}^{\rm x}(z')  \\
\varepsilon_{\rm cl,co}^{\rm x}(z') &0
\end{array}
\right]
\left[
\begin{array}{c}
iG_{\rm co,co}(z',z_i) \\
iG_{\rm cl,co}(z',z_i)
\end{array}
\right],
\label{eq:dyson_num_lpg}
\end{align}
\end{widetext}
where we have set the notation as $\tilde{G}_\textrm{co}(z,z_i)=\tilde{G}_{11}(z,z_i) $ and
$\tilde{G}_\textrm{cl}(z,z_i)= \tilde{G}_{22}(z,z_i)$. The contributions to the total
Green's function from zero to second order in $\hat
\varepsilon^\textrm{x}$ can be seen in Fig.~\ref{fig:lpg}.
Numerically, we can improve the discrete dipole approximation
employed in this paper by rescaling the top and bottom row by
$1/\tilde{G}_\textrm{co}(z,z_i)$ and
$1/\tilde{G}_\textrm{cl}(z,z_i)$, respectively, since the rescaled
total Green's function varies more slowly within the cells than
the original Green's function due to $\Lambda_\textrm{G} \gg
\lambda$ for LPGs. To solve the system we use the
simple iteration method discussed in Sec.~\ref{sec:numeval}. The
iteration may be continued until the desired accuracy is reached.
In this work we iterate until the root mean square of the
residue between two consecutive iterations is less than $10^{-12}$.
The dielectric modulation is chosen as a uniform modulation
$\varepsilon_G (\rrr ) = \Delta \varepsilon (\rrr _\bot
)\frac{1}{2}(1-\cos\big(\frac{2\pi}{\Lambda_\textrm{G}})z\big)$.
We model a typical fiber LPG with resonance tentatively at
$\lambda \simeq 1550{\rm nm}$ with the peak dielectric
change $\Delta \epsilon =
2.5\times 10 ^{-3}$, a grating period of $\Lambda_\textrm{G} =
500{\rm \mu m }$, and a difference in the effective indices of
$n^\textrm{eff}_\textrm{co}-n^\textrm{eff}_\textrm{cl} =
\frac{1550{\rm nm}}{\Lambda_G}$. The difference in the effective
indices is assumed constant with respect to the wavelength. The
elements of the operator $\hat{\varepsilon}$, are given by
$\frac{\varepsilon^\textrm{s}_\textrm{co,co}(\rrr)}{\varepsilon_G
(\rrr)} = 0.60$,
$\frac{\varepsilon^\textrm{s}_\textrm{cl,cl}(\rrr)}{\varepsilon_G
(\rrr)} = 0.20$, and
$\frac{\varepsilon^\textrm{x}_\textrm{co,cl}(\rrr)}{\varepsilon_G
(\rrr)} =
\frac{\varepsilon^\textrm{x}_\textrm{cl,co}(\rrr)}{\varepsilon_G
(\rrr)} =0.10$. Finally, the dieletric constant of silica is
$\varepsilon_\textrm{si} = 2.10$.

With the given parameters we model the spectra of the grating with
four different lengths $L = \frac{\pi}{10\kappa} = 8 {\rm mm} ,
\frac{\pi}{3\kappa}= 13.5 {\rm mm}, \frac{\pi}{2\kappa} = 20 {\rm
mm}, \frac{\pi}{\kappa} = 40 {\rm mm}$ corresponding to 16, 27,
40, and 80 grating periods, respectively, and $\kappa$ is given by
the maximum of Eq.~(\ref{eq:kappa_LPG}) with respect to
wavelength.

\begin{figure}[h]
\begin{center}
\includegraphics[angle = 0, width = 0.5\textwidth]{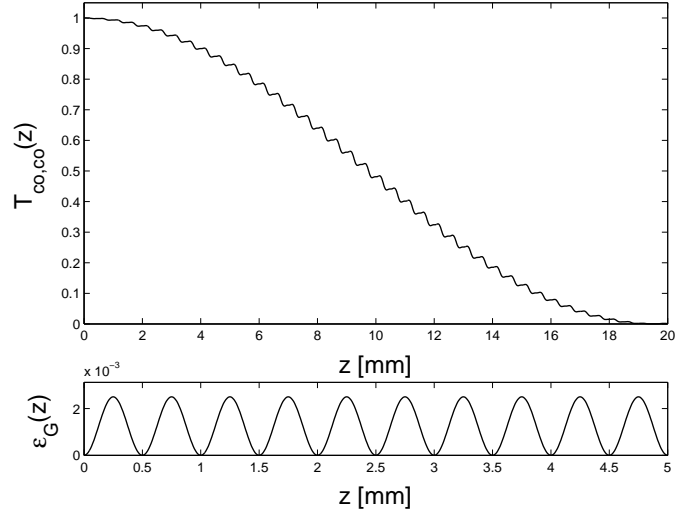}
\caption{Uniform LPG at the resonant wavelength. Top: The
transmission of the core mode, $T_{co}$, as function of the
grating length. Bottom: the uniform dielectric modulation of the
grating.} \label{fig:trans_vs_L}
\end{center}
\end{figure}

\begin{figure}[h]
\begin{center}
\includegraphics[angle = 0, width = 0.5\textwidth]{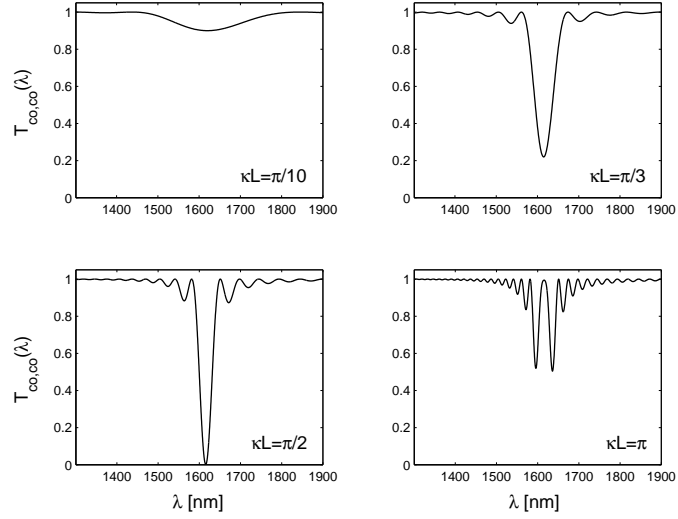}
\caption{Spectra of uniform LPGs with a uniform dielectric
modulation (bottom in Fig.~\ref{fig:trans_vs_L}). The spectra is
plot calculated for four different for grating lengths, $z_f-z_i
= \frac{\pi}{10\kappa}, \frac{\pi}{3\kappa}, \frac{\pi}{2\kappa},
\frac{\pi}{\kappa}$ corresponding to 16, 27, 40, and 80 grating
periods, respectively. At $\kappa L = \frac{\pi}{2}$ (bottom left)
all core mode power is transferred into the cladding mode and back
again into the core mode at $\kappa L = \pi$ (bottom right). }
\label{fig:4kappa}
\end{center}
\end{figure}

\begin{figure}[h]
\begin{center}
\includegraphics[angle = 0, width = 0.5\textwidth]{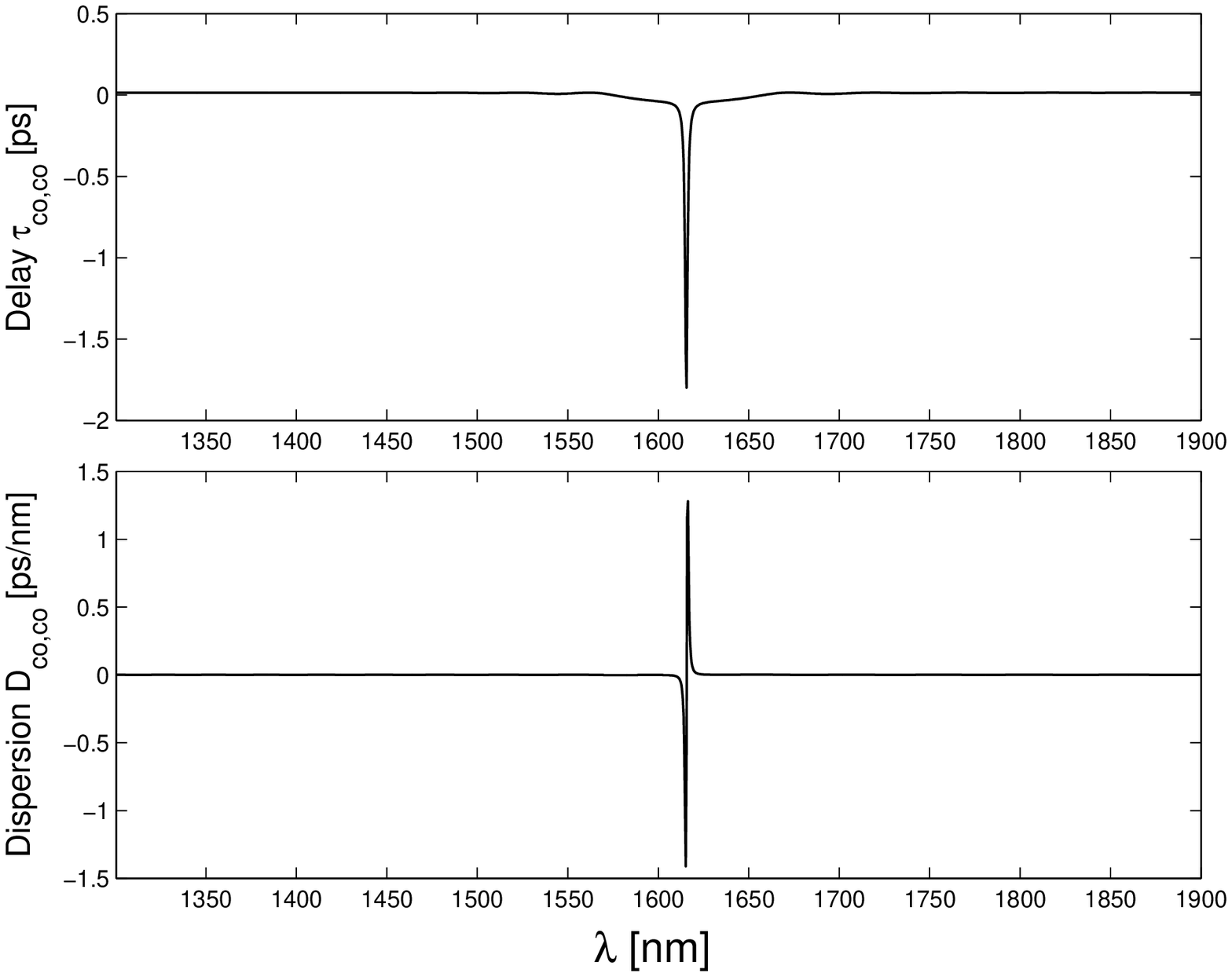}
\caption{Time delay and dispersion of the uniform long-period
grating with length $z_f-z_i  = \frac{\pi}{2\kappa}$. The
corresponding spectrum is seen in the bottom left of
Fig.~\ref{fig:4kappa}.} \label{fig:disp_delay_kappa_pi2}
\end{center}
\end{figure}

\begin{figure}[h]
\begin{center}
\includegraphics[angle = 0, width = 0.5\textwidth]{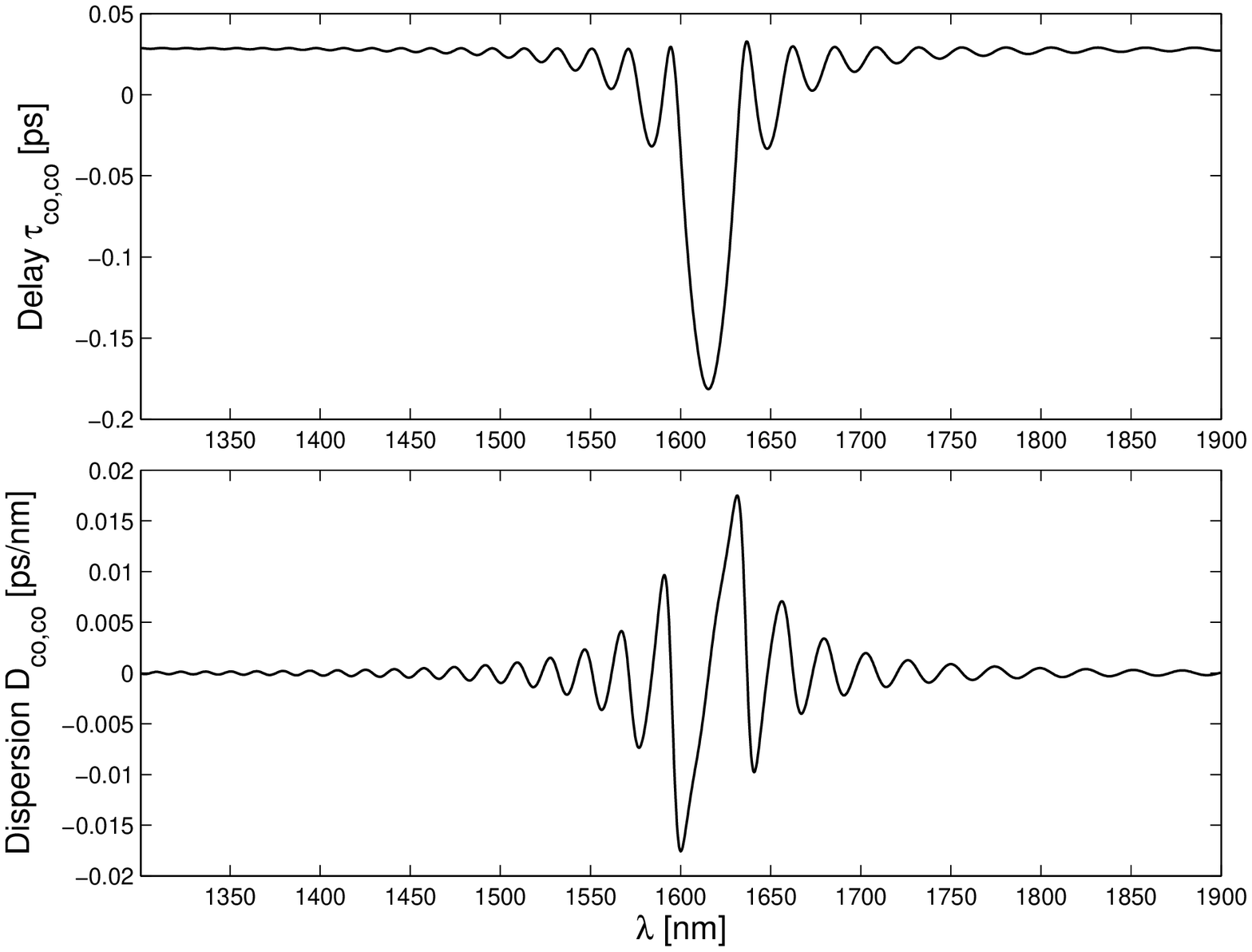}
\caption{Time delay and dispersion of the uniform long-period
grating with length $z_f-z_i  = \frac{\pi}{\kappa}$. The
corresponding spectrum is seen in the bottom right of
Fig.~\ref{fig:4kappa}.} \label{fig:disp_delay_kappa_pi}
\end{center}
\end{figure}

The LPG couples the  core mode power into the cladding mode with
cosine-like amplitude behavior as seen in
Fig.~\ref{fig:trans_vs_L}. In passing we note that the LPG
effectively corresponds to a rotation of the state vector in the
Hilbert space spanned by
$\big\{\EEE_\textrm{co},\EEE_\textrm{cl}\big\}$. The
$\frac{\pi}{2}$ rotation seen in the bottom left figure in
Fig.~\ref{fig:4kappa} can be made into a $\pi$ rotation seen in
the bottom right figure in Fig.~\ref{fig:4kappa} by doubling the
length of the grating. The resonant wavelength found by
Eq.~(\ref{eq:kappa_LPG}) is 1615.7 nm. This includes the
self-coupling of the modes which causes a considerable shift of
the resonant wavelength of 65.7 nm.
The time delay and the dispersion of the core transmission
 can be calculated with
Eqs.~(\ref{eq:groupdelay},\ref{eq:dispersion}) and is seen in
Figs.~\ref{fig:disp_delay_kappa_pi2},~\ref{fig:disp_delay_kappa_pi}.
The LPG has a low group delay and dispersion
making it suitable as a dispersion-free wide-band rejection
filter.

\begin{figure}[h]
\begin{center}
\includegraphics[angle = 0, width = 0.5\textwidth]{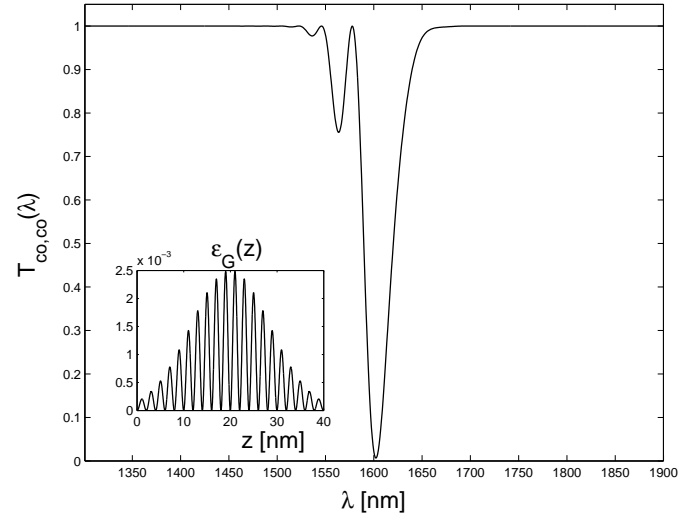}
\caption{Apodised LPG. The side-lobes seen in the bottom left of
Fig.~\ref{fig:4kappa} have been reduced. Insert: a schematic view
of the raised Gaussian apodised dielectric modulation. The grating
period has been exaggerated with respect to the grating length for
clarity.} \label{fig:apodisation}
\end{center}
\end{figure}

\begin{figure}[h]
\begin{center}
\includegraphics[angle = 0, width = 0.5\textwidth]{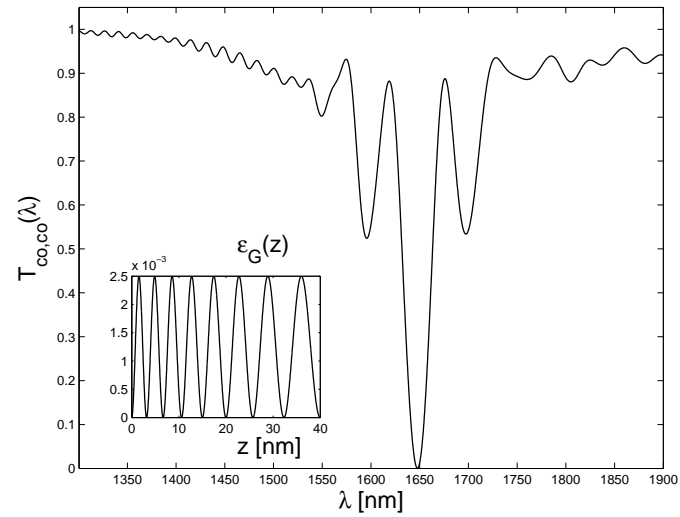}
\caption{Linearly chirped LPG. The side-lobes seen in the bottom
left of Fig.~\ref{fig:4kappa} have been enhanced rather than
reduced. Insert: a schematic view of the linearly chirped grating
period of the dielectric modulation. The grating periods and the
chirp have been exaggerated with respect to the grating length for
clarity.} \label{fig:chirp}
\end{center}
\end{figure}

If the dielectric modulation has a Gaussian envelope the grating
is said to be raised Gaussian apodised~\cite{deyerl2004}. The
so-called taper parameter which determines the strength of the
apodisation is $g = \frac{1}{2}$. In Fig.~\ref{fig:apodisation} is
shown the spectrum for a grating that is 40 mm long with 80
grating periods. The apodisation is seen to reduce the side-lobes
seen in Fig.~\ref{fig:4kappa} We also model a linearly chirped
grating with the chirp constant chosen as $c_1 = 0.005$. Thus the
grating period varies 0.5\% across the length. The number of
grating periods is 40 corresponding to a total grating length of
20 mm and apart from the chirp the resulting spectrum would be
identical to the one seen bottom left in Fig.~\ref{fig:4kappa}.
The spectrum of the chirped grating can be seen in
Fig.~\ref{fig:chirp}. The 0.5\% variation in the grating period is
seen to have a clear effect on the spectrum. The side-lobes of the
same magnitude as those of the uniform grating and the resonant
wavelength is shifted to 1631.6 nm from 1615.7 nm even though the
average grating period has remained the same.
\section{Counter-propagating waves - Bragg gratings}
\label{bragggratings}
As explained in Sec.~\ref{opticalwaveguide} BGs couple the
counter-propagting part of the same mode. The coupling of the
copropagating modes is so far from resonance that it is neglected.
We will in the following suppress the mode subscript.
\subsection{Resonance condition}
Following the same line of arguments as in Sec.~\ref{sec:res_lpg} we arrive at
\begin{align}
\kappa(z',z)\cdot(z'-z) =&
\frac{k^2}{2\beta} \int_{z_i}
^{z_f}\hspace{-0.25cm} dz'' \hspace{-0cm} &e^{-i2\tilde{\beta}
(z'') z''} \varepsilon^x(z'')
\label{eq:kappa_BG}.
\end{align}
This expression resembles Eq.~(\ref{eq:kappa_LPG}) but has
different integration limits due to the step function in
Eq.~(\ref{eq:intermediategreensfunction-}). For a uniform grating
we may again conclude that self-coupling changes the resonance
condition from $2\beta = \frac{2\pi}{\Lambda_G}$ to an average
$2\overline {\beta(z)}  =\frac{2\pi}{\Lambda_G}$ in the limit
$\frac{z_f-z_i}{\Lambda_G}\to \infty$, which also is the result of
the synchronous approximation used in
CMT~\cite{Erdogan:1997,Erdogan:1997IEEE}.

\subsection{Numerical solution of Dyson's equation}
\begin{figure}[h]
\begin{center}
\includegraphics[angle = 0, width =
0.5\textwidth]{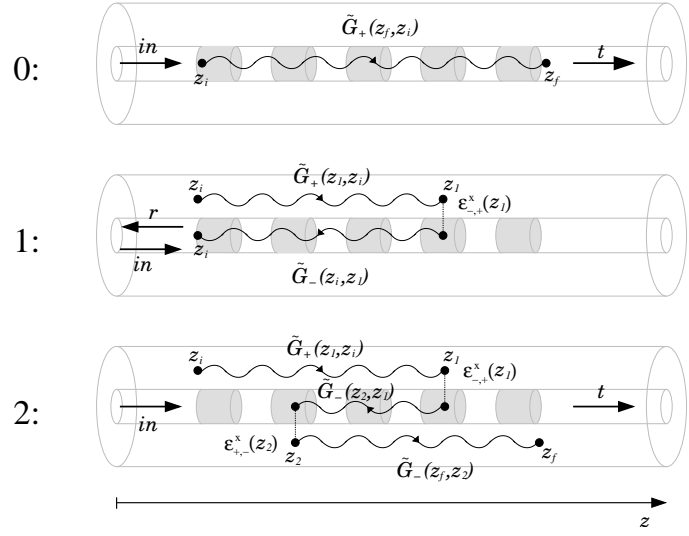} \caption{Diagrammatic representation
of the zero to second order contributions to the total Green's
function Eq.~(\ref{eq:dyson_num_bg}) for a BG. An incident core
mode interacts with  its oppositely propagating counterpart
through the dielectric modulation. } \label{fig:diag_BG}
\end{center}
\end{figure}

We will solve the Dyson's equation (\ref{eq:dysontilde}) by direct
numerical solution. Since we consider a system where
electric mode is incident from the left, we only have to
evaluate the first column of the matrix Dyson's equation
\begin{widetext}
\begin{align}
\left[
\begin{array}{cc}
iG_{\rm +,+}(z,z_i)   \\
iG_{\rm -,+}(z,z_i)
\end{array}
\right]
&=
\left[
\begin{array}{cc}
i\tilde{G}_{\rm +}(z,z_i)  \\
0
\end{array}
\right]
-ik^2
\left[
\begin{array}{cc}
\int_{z_i}^{z}\hspace{-0.cm}dz'
i\tilde{G}_{\rm +}(z,z') & 0 \\
0 &
\int_{z}^{z_f}\hspace{-0.cm}dz'
i\tilde{G}_{\rm -}(z,z')
\end{array}
\right]
\left[
\begin{array}{cc}
0 &\varepsilon_{\rm +,-}^{\rm x}(z')  \\
\varepsilon_{\rm -,+}^{\rm x}(z') &0
\end{array}
\right]
\left[
\begin{array}{c}
iG_{\rm +,+}(z',z_i) \\
iG_{\rm -,+}(z',z_i)
\end{array}
\right],
\label{eq:dyson_num_bg}
\end{align}
\end{widetext}
where we have abbreviated $\tilde{G}_{\rm +} = \tilde G_{\rm +,+}$ and
$\tilde{G}_{\rm -} = \tilde G_{\rm -,-}$. The equation
(\ref{eq:dyson_num_bg}) may be solved in the discrete dipole
approximation by simple iteration as discussed in
Sec.~\ref{sec:numeval}. The expressions obtained by simple
iteration from zero order through second order can seen in
Fig.~\ref{fig:diag_BG} in a diagrammatic representation. It is
seen that a reflected wave has undergone an uneven number of
interactions whereas the transmitted wave has undergone an even
number of interactions. This fact is true for all orders of
interactions.

The iteration may be continued until the desired accuracy is
reached. In this report we iterate until the root mean square of
the residue between two consecutive iterations is less than
$10^{-12}$. When simple iteration of Dyson's equation converges
this may be met in less than 20 iterations. Simple iteration often
diverges for strong BGs. For uniform gratings it is possible to
analyze the divergence. We will not discuss this analysis in
detail here, but only assert that simple iteration converges for
uniform gratings when $|\kappa|L < \pi/2$ implying $T_{co,co}
\gtrsim 0.16$.. For stronger and non-uniform gratings Dyson's
equation (\ref{eq:dyson_num_bg}) may be solved as a
self-consistency problem in  $iG_{\rm +,+}(z',z_i)$. In this paper
we have employed a generalised minimum residue method for stronger
gratings~\cite{mathworks}. In our numerical algorithm the
generalised minimum residue method automatically takes over if
simple iteration of Dyson's equation diverges.

\begin{figure}[h]
\begin{center}
\includegraphics[angle = 0, width = 0.5\textwidth]{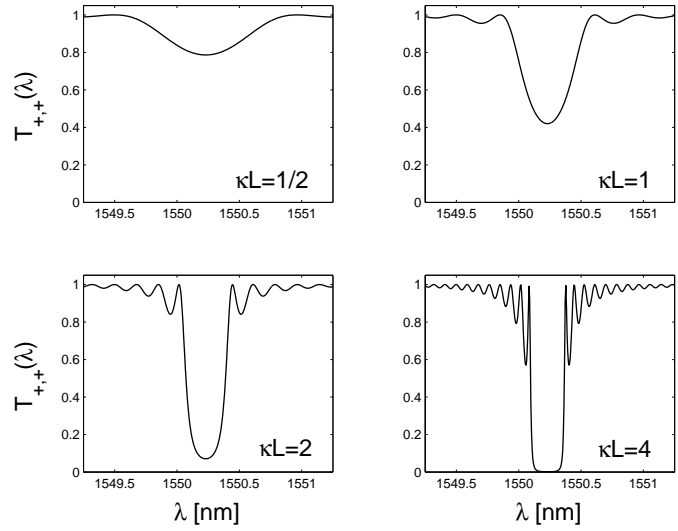}
\caption{Uniform BG spectra for four different lengths, $\kappa L
= 1/2, 1, 2, 4$. The Bragg rejection band is emerging for
increasing $\kappa L$.} \label{fig:BGuni_4}
\end{center}
\end{figure}

\begin{figure}[h]
\begin{center}
\includegraphics[angle = 0, width = 0.5\textwidth]{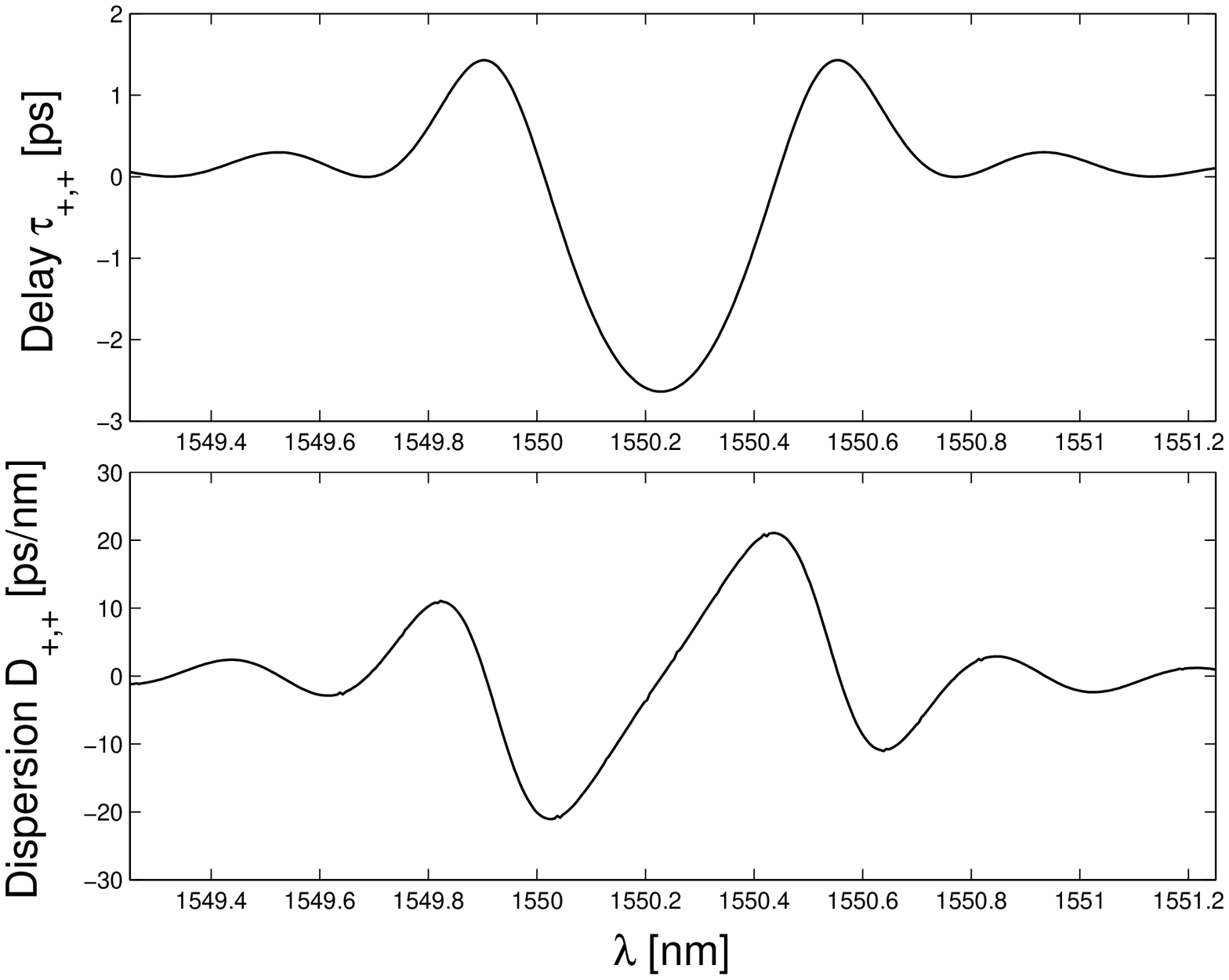}
\caption{Group delay and dispersion for the transmission
coefficient of a uniform BG with $\kappa L = 1$. The corresponding
spectrum is shown top right in Fig.~\ref{fig:BGuni_4}.}
\label{fig:disp_delay_kL=1}
\end{center}
\end{figure}

\begin{figure}[h]
\begin{center}
\includegraphics[angle = 0, width = 0.5\textwidth]{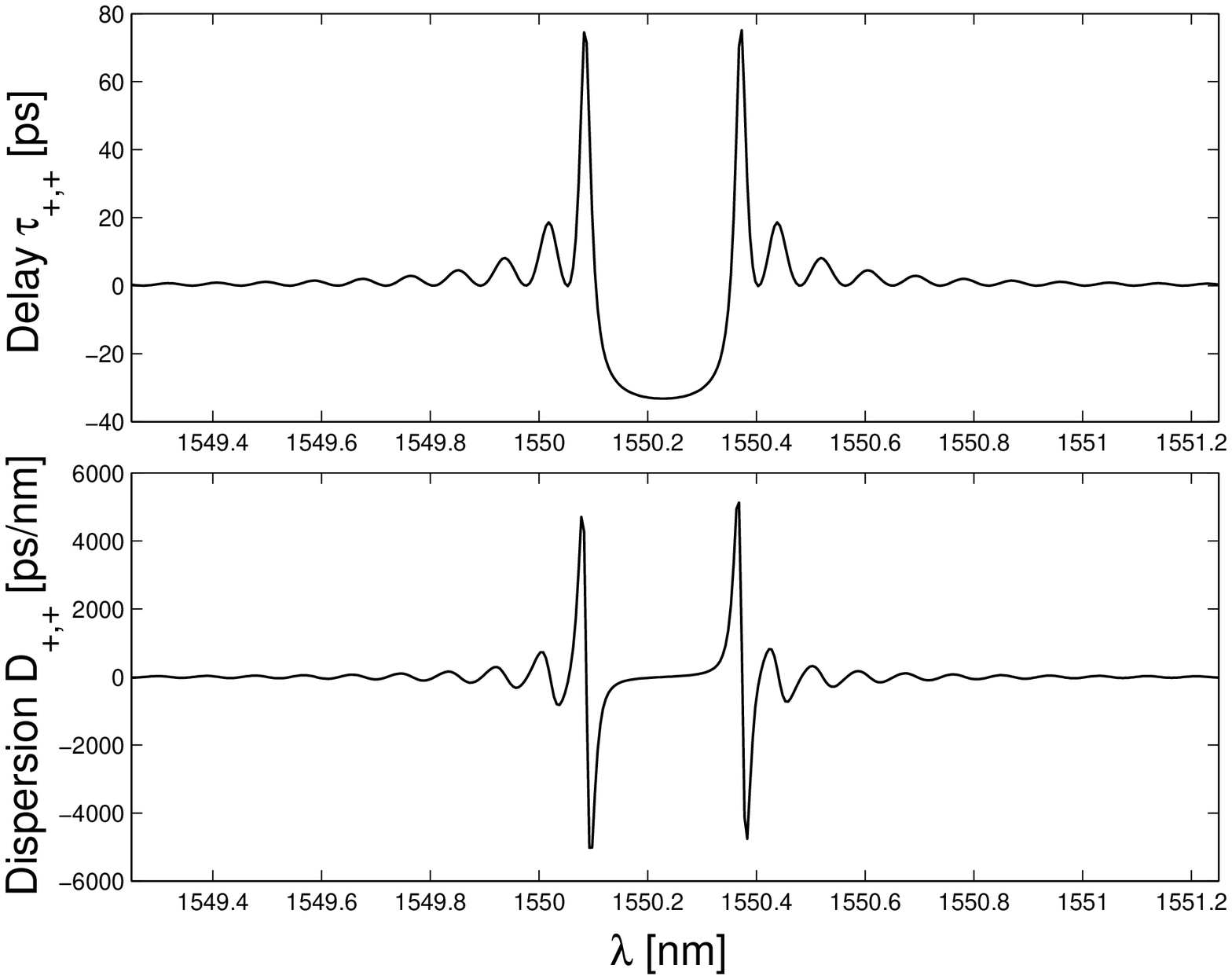}
\caption{Group delay and dispersion for the transmission
coefficient of a uniform BG with $\kappa L = 4$. The corresponding
spectrum is shown top right in Fig.~\ref{fig:BGuni_4}}
\label{fig:disp_delay_kL=4}
\end{center}
\end{figure}

We choose the parameters of the gratings to closely resemble a
standard fiber BG with resonance tentatively at
$\lambda\simeq 1550{\rm nm}$, leading to a grating period of
$\Lambda_G  = \frac{1550{\rm nm}}{2} \frac{k}{\beta}$. The peak
dielectric modulation is $\Delta\epsilon = 2.5\times 10^{-3}$, and
the elements of the dielectric operator $\hat \varepsilon$ are
given by
$\frac{\varepsilon^\textrm{s}_\textrm{+,+}(\rrr)}{\varepsilon_G
(\rrr)} =
\frac{\varepsilon^\textrm{s}_\textrm{-,-}(\rrr)}{\varepsilon_G
(\rrr)}  = \frac{\varepsilon^\textrm{x}_\textrm{+,-}(\rrr)}{
\varepsilon_G (\rrr)}=
\frac{\varepsilon^\textrm{x}_\textrm{-,+}(\rrr)}{ \varepsilon_G
(\rrr)} = 0.50$.
BGs with four different lengths are shown in
Fig.~\ref{fig:BGuni_4}. The length is given in the terms of the
coupling constant at resonance, $\kappa$, given by
Eq.~(\ref{eq:kappa_BG}). The lengths of the four gratings are
1.145mm,  2.290mm,   4.581mm,  9.161mm, for $\kappa L = 1/2, 1, 2,
4$, and with 2143, 4285, 8570, 17140, grating periods,
respectively. Group delay and dispersion of the core transmission
can be found using the
Eqs.~(\ref{eq:groupdelay},\ref{eq:dispersion}). The dispersion and
delay for the transmission coefficient for the $\kappa L = 1$ and
$\kappa L = 4$ are shown in Fig.~\ref{fig:disp_delay_kL=1} and
Fig.~\ref{fig:disp_delay_kL=4}, respectively.

\begin{figure}[h]
\begin{center}
\includegraphics[angle = 0, width = 0.5\textwidth]{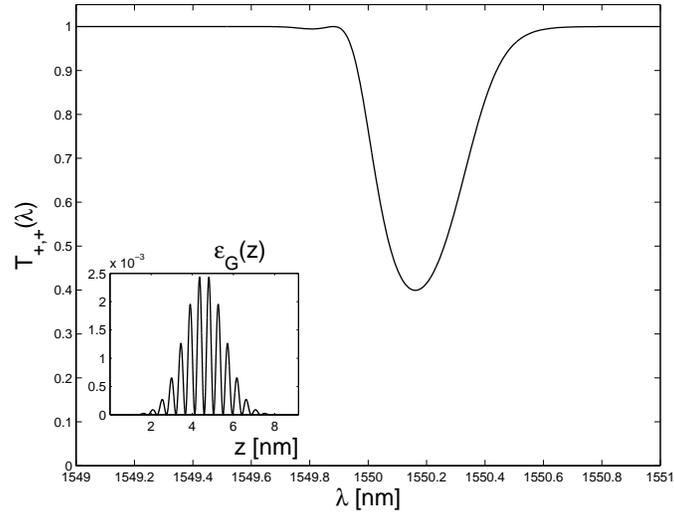}
\caption{Spectrum for a raised-Gaussian apodised BG. Insert: the
raised-Gaussian dielectric modulation. The grating period has been
exaggerated with respect to the length of the grating for
clarity.} \label{fig:BG_apod}
\end{center}
\end{figure}

\begin{figure}[h]
\begin{center}
\includegraphics[angle = 0, width = 0.5\textwidth]{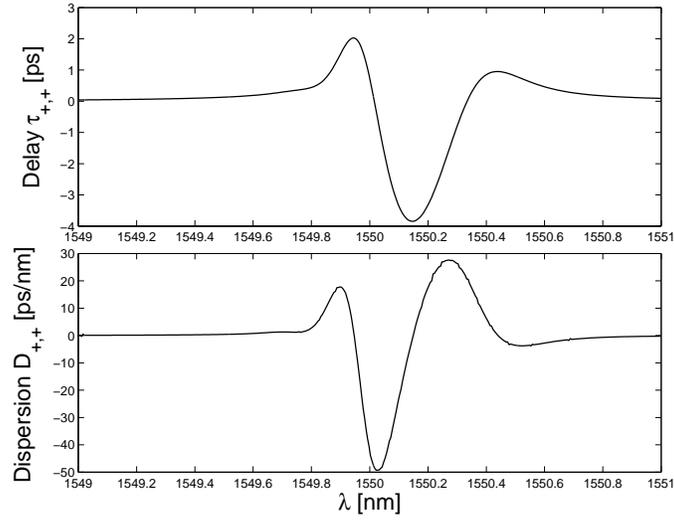}
\caption{Group delay and dispersion for a raised-Gaussian apodised
BG. The spectrum of the grating is shown in
Fig.~\ref{fig:BG_apod}.} \label{fig:BG_apod_delay_disp}
\end{center}
\end{figure}

If we use raised-Gaussian apodisation~\cite{deyerl2004} with
length L = 9.161mm and the taper parameter, $g = 1/4$ we get the
spectrum in Fig.~\ref{fig:BG_apod}. Apart from the apodisation the
grating has the same parameters as the uniform grating shown
bottom right in Fig.~\ref{fig:BGuni_4}. The apodisation has
removed the side lobes and riddles seen in spectrum of the uniform
grating, but has lower reflection (higher transmission) at
resonance. The group delay and dispersion are also smoother for
the apodised grating as seen in Fig.~\ref{fig:BG_apod_delay_disp}.

\begin{figure}[h]
\begin{center}
\includegraphics[angle = 0, width = 0.5\textwidth]{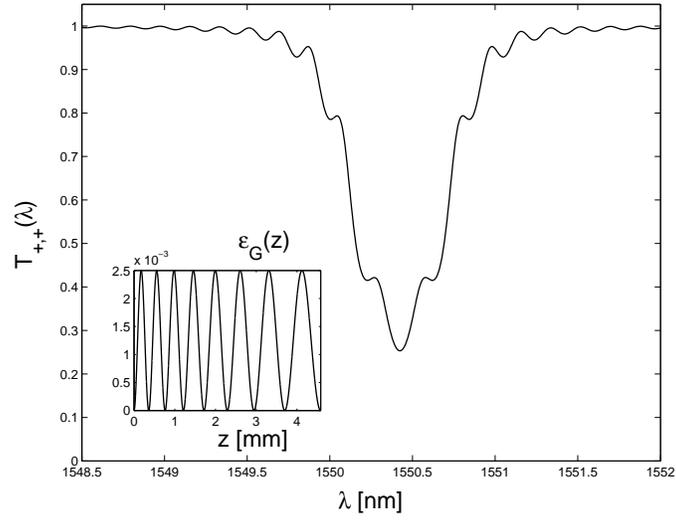}
\caption{Spectrum for a linearly chirped BG. Insert: the linearly
chirped dielectric modulation. The chirp  has been exaggerated
with respect to the period of the grating for clarity. }
\label{fig:BG_chirp}
\end{center}
\end{figure}

\begin{figure}[h]
\begin{center}
\includegraphics[angle = 0, width = 0.5\textwidth]{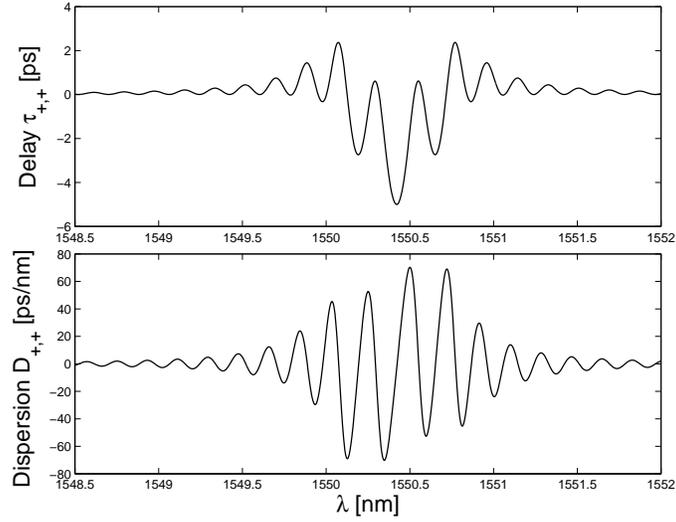}
\caption{Group delay and dispersion for a linearly chirped BG. The
spectrum of the grating is shown in Fig.~\ref{fig:BG_chirp}.}
\label{fig:BG_chirp_delay_disp}
\end{center}
\end{figure}

Linear chirp has been incorporated in the grating seen in
Fig.~\ref{fig:BG_chirp}. The chirp parameter is chosen as $c_1=
5\times10^{-5}$ and the length of the grating is 4.581mm. The
group delay and dispersion of the transmission is seen in
Fig.~\ref{fig:BG_chirp_delay_disp}. It is seen that the chirp
affects both the spectrum and the group delay and dispersion when
comparing with a uniform grating.

\section{Discussion}
\label{discussion}
The numerical results of GFM differ from the widely used CMY in
two respects. The small oscillation seen in
Fig.~\ref{fig:trans_vs_L} does not occur in CMT, since it results
from the second order derivative which is neglected in CMT. In GFM
the oscillation occurs for both LPGs and BGs, but vanishes with
decreasing coupling strength and averages out in general. CMT
relies on a Fourier decomposition of the dielectric modulation.
This is a very good approximation when the grating is uniform and
has a high number of periods. An LPG usually have a modest number
of periods compared with an BG, and the GFM spectra differ from
CMT spectra in the side-lobes, while they coincide around the
resonance. The GFM could therefore be applied for better
apodisation of LPGs. BGs have a large number of periods, and thus
CMT and GFM agree well for the uniform and the weakly apodised and
chirped BGs presented in this report. However, BGs are being
designed with increasing complex dielectric modulations to obtain
extraordinary spectral and dispersive characteristics. The GFM may
here be useful in the respect,  that it does not divide the
dielectric modulation into uniform segments, and therefore offers
more freedom when designing the apodised dielectric modulation.
The GFM could also be used to study the mechanisms of apodisation,
which are still not completely understood \cite{minzioni2006}.

 The examples considered have all been fiber
gratings. The GFM method is exact for transverse electric modes,
but some optical fibers do not support these transverse modes.
However the non-transverse part of a mode usually only holds a
very small fraction of the total electrical field and contributes
only negligible to the grating characteristics.

Although the presented theory is exact from the transverse
electrical wave equation Eq.~(\ref{eq:eigenfreq_r}) the theory is
in fact only exact within the basis set of the electric fields of
the reference system (in finite element methods this is known as
Galerkin error being zero). The electrical fields of the reference
system may or may not describe the grating well. This is actually
a feature in all grating theories and should not cause any
worries, but some consideration should be taken when applying the
GFM to high-index contrast systems. A discussion of the
variational error of the basis set can be found
in~\cite{rindorf2006a}.

The Green's functions and the Dyson's equation have here been used
to model linear dielectric modulations. The theory can be adapted to
a non-linear dielectric modulations, both local and non-local, for
simulating harmonic generation, soliton dynamics and other
non-linear effects. If a non-local dielectric modulation is used then powerful
perturbative methods can be employed such as the Feynman diagrams
used in many-body physics.

In Figs.~\ref{fig:disp_delay_kL=4},~\ref{fig:disp_delay_kappa_pi2}
it is seen that the group delay for both Bragg and long-period
gratings is negative near resonance. This may seem conflicting
with casuality, but physically it is perfectly possible that the
group velocity (pulse envelope velocity) is greater than the speed
of light or even zero~\cite{mitchell1999} causing negative time
delays. However, numerical simulations indicate that negative
group velocities are impossible in passive one-dimensional
periodic media~\cite{poirier2005}.

\section{Conclusion}
\label{conclusion}
A method for calculation of optical-waveguide gratings' optical
characteristics based on Green's functions and the Dyson's
equation has been presented. We have presented accurate  spectra
for long-period and both uniform and apodised BGs as well as the
group delay and dispersion characteristics.

The GFM is exact for transverse electric modes for gratings with
arbitrary dielectric modulations in both the long-period and the
BG application. The method incorporates self-coupling
in an exact analytic manner and also gives an exact resonance
condition for gratings with arbitrary dielectric modulations.  The
GFM relies on iterative solution of integral equations which
scales as $\mathcal{O}(N)$ which may be solved to within any given
accuracy.

The GFM may also be useful in simulating non-linear problems or
imperfect gratings with random mode scattering.

\bibliographystyle{apsrev}

\begin{thebibliography}{37}
\expandafter\ifx\csname natexlab\endcsname\relax\def\natexlab#1{#1}\fi
\expandafter\ifx\csname bibnamefont\endcsname\relax
  \def\bibnamefont#1{#1}\fi
\expandafter\ifx\csname bibfnamefont\endcsname\relax
  \def\bibfnamefont#1{#1}\fi
\expandafter\ifx\csname citenamefont\endcsname\relax
  \def\citenamefont#1{#1}\fi
\expandafter\ifx\csname url\endcsname\relax
  \def\url#1{\texttt{#1}}\fi
\expandafter\ifx\csname urlprefix\endcsname\relax\def\urlprefix{URL }\fi
\providecommand{\bibinfo}[2]{#2}
\providecommand{\eprint}[2][]{\url{#2}}

\bibitem[{\citenamefont{Meltz et~al.}(1989)\citenamefont{Meltz, Morey, and
  Glenn}}]{meltz1989}
\bibinfo{author}{\bibfnamefont{G.}~\bibnamefont{Meltz}},
  \bibinfo{author}{\bibfnamefont{W.~W.} \bibnamefont{Morey}}, \bibnamefont{and}
  \bibinfo{author}{\bibfnamefont{W.~H.} \bibnamefont{Glenn}},
  \bibinfo{journal}{Opt. Lett.} \textbf{\bibinfo{volume}{14 (15)}},
  \bibinfo{pages}{823} (\bibinfo{year}{1989}).

\bibitem[{\citenamefont{Hill and Meltz}(1997)}]{hill1997}
\bibinfo{author}{\bibfnamefont{K.~O.} \bibnamefont{Hill}} \bibnamefont{and}
  \bibinfo{author}{\bibfnamefont{G.}~\bibnamefont{Meltz}}, \bibinfo{journal}{J.
  Lightwave Technol.} \textbf{\bibinfo{volume}{15 (8)}}, \bibinfo{pages}{1263}
  (\bibinfo{year}{1997}).

\bibitem[{\citenamefont{Kogelnik and Shank}(1972)}]{kogelnik1972}
\bibinfo{author}{\bibfnamefont{H.}~\bibnamefont{Kogelnik}} \bibnamefont{and}
  \bibinfo{author}{\bibfnamefont{C.}~\bibnamefont{Shank}}, \bibinfo{journal}{J.
  Appl. Phys.} \textbf{\bibinfo{volume}{43}}, \bibinfo{pages}{2327}
  (\bibinfo{year}{1972}).

\bibitem[{\citenamefont{Yariv}(1973)}]{yariv1973}
\bibinfo{author}{\bibfnamefont{A.}~\bibnamefont{Yariv}}, \bibinfo{journal}{IEEE
  J. Quant. Electron.} \textbf{\bibinfo{volume}{QE-9}}, \bibinfo{pages}{919 }
  (\bibinfo{year}{1973}).

\bibitem[{\citenamefont{Erdogan}(1997{\natexlab{a}})}]{Erdogan:1997}
\bibinfo{author}{\bibfnamefont{T.}~\bibnamefont{Erdogan}}, \bibinfo{journal}{J.
  Opt. Soc. Am. A} \textbf{\bibinfo{volume}{14}}, \bibinfo{pages}{1760 }
  (\bibinfo{year}{1997}{\natexlab{a}}).

\bibitem[{\citenamefont{Erdogan}(1997{\natexlab{b}})}]{Erdogan:1997IEEE}
\bibinfo{author}{\bibfnamefont{T.}~\bibnamefont{Erdogan}}, \bibinfo{journal}{J.
  Lightwave Technol.} \textbf{\bibinfo{volume}{15}}, \bibinfo{pages}{1277 }
  (\bibinfo{year}{1997}{\natexlab{b}}).

\bibitem[{\citenamefont{Sipe et~al.}(1994)\citenamefont{Sipe, Poladian, and
  de~Sterke}}]{sipe1994}
\bibinfo{author}{\bibfnamefont{J.~E.} \bibnamefont{Sipe}},
  \bibinfo{author}{\bibfnamefont{L.}~\bibnamefont{Poladian}}, \bibnamefont{and}
  \bibinfo{author}{\bibfnamefont{M.}~\bibnamefont{de~Sterke}},
  \bibinfo{journal}{J. Opt. Soc. Am. A} \textbf{\bibinfo{volume}{11}},
  \bibinfo{pages}{1307} (\bibinfo{year}{1994}).

\bibitem[{\citenamefont{Sullivan and Hall}(1995)}]{sullivan1995}
\bibinfo{author}{\bibfnamefont{K.~G.} \bibnamefont{Sullivan}} \bibnamefont{and}
  \bibinfo{author}{\bibfnamefont{D.~G.} \bibnamefont{Hall}},
  \bibinfo{journal}{Opt. Commun.} \textbf{\bibinfo{volume}{118}},
  \bibinfo{pages}{509 } (\bibinfo{year}{1995}).

\bibitem[{\citenamefont{Poladian}(1996)}]{poladian1996}
\bibinfo{author}{\bibfnamefont{L.}~\bibnamefont{Poladian}},
  \bibinfo{journal}{Phys. Rev. E} \textbf{\bibinfo{volume}{54}},
  \bibinfo{pages}{2963 } (\bibinfo{year}{1996}).

\bibitem[{\citenamefont{Passaro
  et~al.}(2002{\natexlab{a}})\citenamefont{Passaro, Diana, and
  Armenise}}]{Passaro:2002}
\bibinfo{author}{\bibfnamefont{V.~M.~N.} \bibnamefont{Passaro}},
  \bibinfo{author}{\bibfnamefont{R.}~\bibnamefont{Diana}}, \bibnamefont{and}
  \bibinfo{author}{\bibfnamefont{M.~N.} \bibnamefont{Armenise}},
  \bibinfo{journal}{J. Opt. Soc. Am. A} \textbf{\bibinfo{volume}{19}},
  \bibinfo{pages}{1844 } (\bibinfo{year}{2002}{\natexlab{a}}).

\bibitem[{\citenamefont{Passaro
  et~al.}(2002{\natexlab{b}})\citenamefont{Passaro, Diana, and
  Armenise}}]{Passaro:2002a}
\bibinfo{author}{\bibfnamefont{V.~M.~N.} \bibnamefont{Passaro}},
  \bibinfo{author}{\bibfnamefont{R.}~\bibnamefont{Diana}}, \bibnamefont{and}
  \bibinfo{author}{\bibfnamefont{M.~N.} \bibnamefont{Armenise}},
  \bibinfo{journal}{J. Opt. Soc. Am. A} \textbf{\bibinfo{volume}{19}},
  \bibinfo{pages}{1855 } (\bibinfo{year}{2002}{\natexlab{b}}).

\bibitem[{\citenamefont{Wellerbrophy and Hall}(1985)}]{wellerbrophy1985}
\bibinfo{author}{\bibfnamefont{L.~A.} \bibnamefont{Wellerbrophy}}
  \bibnamefont{and} \bibinfo{author}{\bibfnamefont{D.~G.} \bibnamefont{Hall}},
  \bibinfo{journal}{J. Opt. Soc. Am. A} \textbf{\bibinfo{volume}{2}},
  \bibinfo{pages}{863} (\bibinfo{year}{1985}).

\bibitem[{\citenamefont{Peral et~al.}(1996{\natexlab{a}})\citenamefont{Peral,
  Capmany, and Martin}}]{peral1996a}
\bibinfo{author}{\bibfnamefont{E.}~\bibnamefont{Peral}},
  \bibinfo{author}{\bibfnamefont{J.}~\bibnamefont{Capmany}}, \bibnamefont{and}
  \bibinfo{author}{\bibfnamefont{J.}~\bibnamefont{Marti}},
  \bibinfo{journal}{Electron. Lett.} \textbf{\bibinfo{volume}{32}},
  \bibinfo{pages}{918} (\bibinfo{year}{1996}{\natexlab{a}}).

\bibitem[{\citenamefont{Peral et~al.}(1996{\natexlab{b}})\citenamefont{Peral,
  Capmany, and Martin}}]{peral1996b}
\bibinfo{author}{\bibfnamefont{E.}~\bibnamefont{Peral}},
  \bibinfo{author}{\bibfnamefont{J.}~\bibnamefont{Capmany}}, \bibnamefont{and}
  \bibinfo{author}{\bibfnamefont{J.}~\bibnamefont{Martin}},
  \bibinfo{journal}{IEEE J. Quantum Electron.} \textbf{\bibinfo{volume}{32}},
  \bibinfo{pages}{2078} (\bibinfo{year}{1996}{\natexlab{b}}).

\bibitem[{\citenamefont{Feced et~al.}(1999)\citenamefont{Feced, Zervas, and
  Muriel}}]{feced1999}
\bibinfo{author}{\bibfnamefont{R.}~\bibnamefont{Feced}},
  \bibinfo{author}{\bibfnamefont{M.~N.} \bibnamefont{Zervas}},
  \bibnamefont{and} \bibinfo{author}{\bibfnamefont{M.~A.}
  \bibnamefont{Muriel}}, \bibinfo{journal}{IEEE J. Quantum Electron.}
  \textbf{\bibinfo{volume}{35}}, \bibinfo{pages}{1105} (\bibinfo{year}{1999}).

\bibitem[{\citenamefont{Yamada and Sakuda}(1987)}]{yamada1987}
\bibinfo{author}{\bibfnamefont{M.}~\bibnamefont{Yamada}} \bibnamefont{and}
  \bibinfo{author}{\bibfnamefont{K.}~\bibnamefont{Sakuda}},
  \bibinfo{journal}{Appl. Opt.} \textbf{\bibinfo{volume}{26}},
  \bibinfo{pages}{3474} (\bibinfo{year}{1987}).

\bibitem[{\citenamefont{Bruckstein et~al.}(1985)\citenamefont{Bruckstein, Levy,
  and Kailath}}]{bruckstein1985}
\bibinfo{author}{\bibfnamefont{A.}~\bibnamefont{Bruckstein}},
  \bibinfo{author}{\bibfnamefont{B.}~\bibnamefont{Levy}}, \bibnamefont{and}
  \bibinfo{author}{\bibfnamefont{T.}~\bibnamefont{Kailath}},
  \bibinfo{journal}{SIAM Journal on Applied Mathematics}
  \textbf{\bibinfo{volume}{45}}, \bibinfo{pages}{312} (\bibinfo{year}{1985}).

\bibitem[{\citenamefont{Fetter and Walecka}(1971)}]{fetterwalecka}
\bibinfo{author}{\bibfnamefont{A.~L.} \bibnamefont{Fetter}} \bibnamefont{and}
  \bibinfo{author}{\bibfnamefont{J.~D.} \bibnamefont{Walecka}},
  \emph{\bibinfo{title}{Quantum Theory of Many-Particle Systems}}
  (\bibinfo{publisher}{McGraw-Hill, Inc.}, \bibinfo{year}{1971}).

\bibitem[{\citenamefont{Bruus and Flensberg}(2004)}]{Bruus}
\bibinfo{author}{\bibfnamefont{H.}~\bibnamefont{Bruus}} \bibnamefont{and}
  \bibinfo{author}{\bibfnamefont{K.}~\bibnamefont{Flensberg}},
  \emph{\bibinfo{title}{Many-Body Quantum Theory in Condensed Matter Physics}}
  (\bibinfo{publisher}{Oxford University Press}, \bibinfo{address}{Oxford},
  \bibinfo{year}{2004}).

\bibitem[{\citenamefont{Martin et~al.}(1995)\citenamefont{Martin, Girard, and
  Dereux}}]{martin1995}
\bibinfo{author}{\bibfnamefont{O.~J.~F.} \bibnamefont{Martin}},
  \bibinfo{author}{\bibfnamefont{C.}~\bibnamefont{Girard}}, \bibnamefont{and}
  \bibinfo{author}{\bibfnamefont{A.}~\bibnamefont{Dereux}},
  \bibinfo{journal}{Phys. Rev. Lett.} \textbf{\bibinfo{volume}{74}},
  \bibinfo{pages}{526} (\bibinfo{year}{1995}).

\bibitem[{\citenamefont{Martin et~al.}(1999)\citenamefont{Martin, Girard,
  Smith, and Schultz}}]{martin1999}
\bibinfo{author}{\bibfnamefont{O.~J.~F.} \bibnamefont{Martin}},
  \bibinfo{author}{\bibfnamefont{C.}~\bibnamefont{Girard}},
  \bibinfo{author}{\bibfnamefont{D.~R.} \bibnamefont{Smith}}, \bibnamefont{and}
  \bibinfo{author}{\bibfnamefont{S.}~\bibnamefont{Schultz}},
  \bibinfo{journal}{Phys. Rev. Lett.} \textbf{\bibinfo{volume}{82}},
  \bibinfo{pages}{315} (\bibinfo{year}{1999}).

\bibitem[{\citenamefont{Rahachou and Zozoulenko}(2005)}]{rahachou2005}
\bibinfo{author}{\bibfnamefont{A.~I.} \bibnamefont{Rahachou}} \bibnamefont{and}
  \bibinfo{author}{\bibfnamefont{I.~V.} \bibnamefont{Zozoulenko}},
  \bibinfo{journal}{Phys. Rev. B}
  \textbf{\bibinfo{volume}{72}}, \bibinfo{eid}{155117}
(\bibinfo{year}{2005}).

\bibitem[{\citenamefont{Zhao et~al.}(2005)\citenamefont{Zhao, Wang, Gu, and
  Yang}}]{zhao2005}
\bibinfo{author}{\bibfnamefont{L.-M.} \bibnamefont{Zhao}},
  \bibinfo{author}{\bibfnamefont{X.-H.} \bibnamefont{Wang}},
  \bibinfo{author}{\bibfnamefont{B.-Y.} \bibnamefont{Gu}}, \bibnamefont{and}
  \bibinfo{author}{\bibfnamefont{G.-Z.} \bibnamefont{Yang}},
  \bibinfo{journal}{Phys. Rev E} \textbf{\bibinfo{volume}{72}}, \bibinfo{eid}{026614}
       (\bibinfo{year}{2005}).

\bibitem[{\citenamefont{Cross and Kogelnik}(1977)}]{cross1977}
\bibinfo{author}{\bibfnamefont{P.~S.} \bibnamefont{Cross}} \bibnamefont{and}
  \bibinfo{author}{\bibfnamefont{H.}~\bibnamefont{Kogelnik}},
  \bibinfo{journal}{Opt. Lett.} \textbf{\bibinfo{volume}{1}},
  \bibinfo{pages}{43} (\bibinfo{year}{1977}).

\bibitem[{\citenamefont{Ibsen et~al.}(1998)\citenamefont{Ibsen, Durkin, Cole,
  and Laming}}]{ibsen1998}
\bibinfo{author}{\bibfnamefont{M.}~\bibnamefont{Ibsen}},
  \bibinfo{author}{\bibfnamefont{M.}~\bibnamefont{Durkin}},
  \bibinfo{author}{\bibfnamefont{M.}~\bibnamefont{Cole}}, \bibnamefont{and}
  \bibinfo{author}{\bibfnamefont{R.}~\bibnamefont{Laming}},
  \bibinfo{journal}{IEEE Photonic. Technol. Lett.}
  \textbf{\bibinfo{volume}{10}}, \bibinfo{pages}{842 } (\bibinfo{year}{1998}).

\bibitem[{\citenamefont{Ouellete}(1987)}]{ouellete1987}
\bibinfo{author}{\bibfnamefont{F.}~\bibnamefont{Ouellete}},
  \bibinfo{journal}{Opt. Lett.} \textbf{\bibinfo{volume}{12}},
  \bibinfo{pages}{847} (\bibinfo{year}{1987}).

\bibitem[{\citenamefont{Roman and Winick}(1993)}]{roman1993}
\bibinfo{author}{\bibfnamefont{J.~E.} \bibnamefont{Roman}} \bibnamefont{and}
  \bibinfo{author}{\bibfnamefont{K.~A.} \bibnamefont{Winick}},
  \bibinfo{journal}{IEEE J. Quantum Electron.} \textbf{\bibinfo{volume}{29}},
  \bibinfo{pages}{975} (\bibinfo{year}{1993}).

\bibitem[{\citenamefont{Lee}(2003)}]{lee2003}
\bibinfo{author}{\bibfnamefont{B.}~\bibnamefont{Lee}},
  \bibinfo{journal}{Optical Fiber Technology} \textbf{\bibinfo{volume}{9}},
  \bibinfo{pages}{57} (\bibinfo{year}{2003}).

\bibitem[{\citenamefont{Deyerl et~al.}(2004)\citenamefont{Deyerl, Plougmann,
  Jensen, Floreani, Sorensen, and Kristensen}}]{deyerl2004}
\bibinfo{author}{\bibfnamefont{H.~J.} \bibnamefont{Deyerl}},
  \bibinfo{author}{\bibfnamefont{N.}~\bibnamefont{Plougmann}},
  \bibinfo{author}{\bibfnamefont{J.~B.} \bibnamefont{Jensen}},
  \bibinfo{author}{\bibfnamefont{F.}~\bibnamefont{Floreani}},
  \bibinfo{author}{\bibfnamefont{H.~R.} \bibnamefont{Sorensen}},
  \bibnamefont{and}
  \bibinfo{author}{\bibfnamefont{M.}~\bibnamefont{Kristensen}},
  \bibinfo{journal}{Appl. Opt.} \textbf{\bibinfo{volume}{43 (17)}},
  \bibinfo{pages}{3513} (\bibinfo{year}{2004}).

\bibitem[{\citenamefont{Joannopoulos et~al.}(1995)\citenamefont{Joannopoulos,
  Meade, and Winn}}]{joannopoulos}
\bibinfo{author}{\bibfnamefont{J.~D.} \bibnamefont{Joannopoulos}},
  \bibinfo{author}{\bibfnamefont{R.~D.} \bibnamefont{Meade}}, \bibnamefont{and}
  \bibinfo{author}{\bibfnamefont{J.~N.} \bibnamefont{Winn}},
  \emph{\bibinfo{title}{Photonic crystals: molding the flow of light}}
  (\bibinfo{publisher}{Princeton University Press},
  \bibinfo{address}{Princeton}, \bibinfo{year}{1995}).

\bibitem[{\citenamefont{Draine and Flatau}(1994)}]{draine1994}
\bibinfo{author}{\bibfnamefont{B.~T.} \bibnamefont{Draine}} \bibnamefont{and}
  \bibinfo{author}{\bibfnamefont{P.~J.} \bibnamefont{Flatau}},
  \bibinfo{journal}{J. Opt. Soc. Am. A} \textbf{\bibinfo{volume}{11}},
  \bibinfo{pages}{1491} (\bibinfo{year}{1994}).

\bibitem[{mat()}]{mathworks}
\urlprefix\url{http://www.mathworks.com/}.

\bibitem[{\citenamefont{Mitchell and Chiao}(1998)}]{mitchell1999}
\bibinfo{author}{\bibfnamefont{M.}~\bibnamefont{Mitchell}} \bibnamefont{and}
  \bibinfo{author}{\bibfnamefont{R.~Y.} \bibnamefont{Chiao}},
  \bibinfo{journal}{American Journal of Physics} \textbf{\bibinfo{volume}{66}},
  \bibinfo{pages}{14} (\bibinfo{year}{1998}).

\bibitem[{\citenamefont{Poirier et~al.}(2005)\citenamefont{Poirier, Thompson,
  and Hach\'e}}]{poirier2005}
\bibinfo{author}{\bibfnamefont{L.}~\bibnamefont{Poirier}},
  \bibinfo{author}{\bibfnamefont{R.~I.} \bibnamefont{Thompson}},
  \bibnamefont{and} \bibinfo{author}{\bibfnamefont{A.}~\bibnamefont{Hach\'e}},
  \bibinfo{journal}{Opt. Com.} \textbf{\bibinfo{volume}{250}},
  \bibinfo{pages}{258} (\bibinfo{year}{2005}).


\bibitem[{\citenamefont{Minzioni and Tormen}(2006)}]{minzioni2006}
\bibinfo{author}{\bibfnamefont{P.}~\bibnamefont{Minzioni}} \bibnamefont{and}
  \bibinfo{author}{\bibfnamefont{M.} \bibnamefont{Tormen}},
  \bibinfo{journal}{J.  Lightwave Technol.} \textbf{\bibinfo{volume}{24}}, \bibinfo{pages}{605 }
  (\bibinfo{year}{2006}{\natexlab{b}}).

\bibitem[{\citenamefont{Rindorf and Mortensen}(2006)}]{rindorf2006a}
\bibinfo{author}{\bibfnamefont{L.}~\bibnamefont{Rindorf}} \bibnamefont{and}
  \bibinfo{author}{\bibfnamefont{N.~A.} \bibnamefont{Mortensen}},
  \bibinfo{journal}{Opt. Com.} \textbf{\bibinfo{volume}{261}},
  \bibinfo{pages}{181} (\bibinfo{year}{2006}).

\end{thebibliography}

\widetext

\newpage

\end{document}